\documentclass[pra,twocolumn,floatfix,aps,superscriptaddress,amsfonts]{revtex4-1}
\usepackage{graphicx}
\usepackage{array}
\usepackage{amsthm}
\usepackage{times}
\usepackage{bm}
\usepackage{amsmath}
\usepackage[colorlinks,bookmarks=false,citecolor=blue,linkcolor=red,urlcolor=blue]{hyperref}
\usepackage{amssymb}

\begin{document}
\title{One-way deficit and quantum phase transitions in \emph{XY} model and  extended Ising model}

\author{Yao-Kun Wang}
\affiliation{College of Mathematics,  Tonghua Normal University,
 Tonghua, Jilin 134001, China}
\affiliation{Beijing National Laboratory for Condensed Matter Physics, Institute of Physics, Chinese Academy of Sciences, Beijing 100190, China}

\author{Yu-Ran Zhang}
\email{yrzhang@csrc.ac.cn}
\affiliation{Quantum Physics and Quantum Information Division, 
Beijing Computational Science Research Center, Beijing 100094, China}
\affiliation{QCMRG, CEMS, RIKEN, Saitama 351-0198, Japan}
\affiliation{Beijing National Laboratory for Condensed Matter Physics, Institute of Physics, Chinese Academy of Sciences, Beijing 100190, China}

\author{Heng Fan}
%\email{hfan@iphy.ac.cn}
\affiliation{Beijing National Laboratory for Condensed Matter Physics, Institute of Physics, Chinese Academy of Sciences, Beijing 100190, China}
\affiliation{Collaborative Innovation Center of Quantum Matter, Beijing 100190, China}

\begin{abstract}
Originating in questions regarding work extraction from quantum systems coupled to a 
 heat bath, quantum deficit, a kind of quantum correlations besides 
 entanglement and quantum discord, links quantum thermodynamics with quantum correlations.
In this paper, we evaluate the one-way deficit of two adjacent spins in the bulk for the $XY$ model and
its extend model: the extended Ising model. We find that the one-way deficit susceptibility is able to 
characterize the quantum phase transitions in the $XY$ model and even the topological phase transitions 
in the extend Ising model. This study may enlighten extensive studies of quantum phase transitions from 
the perspective of quantum information processing and quantum computation, including finite-temperature phase
transitions, topological phase transitions and dynamical phase transitions of a variety of quantum 
many-body systems.
\end{abstract}
%\eid{identifier}
%\pacs{}
\maketitle

\section{Introduction}

Quantum deficit \cite{oppenheim,horodecki,modi2} is a kind of nonclassical correlation besides 
entanglement and quantum discord. It originates on asking how to use nonlocal operation to 
extract work from a correlated system coupled to a heat bath only in the case of pure states 
\cite{oppenheim}. In the general case, the advantage is related to more general forms of quantum 
correlations. Oppenheim \emph{et al.} defined the work deficit \cite{oppenheim}
%\begin{eqnarray}
%\Delta\equiv W_{t}-W_{l},
%\end{eqnarray}
%where $W_{t}$
as a measure of the difference between the information of the whole system and the localizable 
information \cite{horodecki2,ho03}. Recently, Streltsov \emph{et al.} \cite{Streltsov0,chuan} give the 
definition of the one-way information deficit by means of relative entropy, which is also called one-way 
deficit that uncovers an important role of quantum deficit as a resource for the distribution of entanglement.

Many developments in quantum information processing \cite{key-1} has provided much insight 
into quantum phase transitions in many-body systems\cite{key-2}. Especially, quantum correlations has 
been successful in characterizing a large number of critical phenomena of great interest. In particular, 
entanglement  was the first and most outstanding member for the detection of critical points, see 
Refs.~\cite{key-2,Osterloh,key-3,key-4,Vidal,Osborne}. Furthermore, quantum discord, a significant 
quantum correlation, is also useful for the study on quantum phase transitions \cite{key-8,xzzhang}. Other indications 
of quantumness of ground states of critical systems are also found effective for probing quantum phases 
and quantum phase transitions \cite{key-10,key-11,key-12,key-13,srr,jinjun} and even the identification of 
topological phase transitions \cite{Wen1,Wen2,te1,te2,es,Kitaev,xzzhang,yrzhang}.

In this paper, we calculate the one-way deficit of the thermal ground states of two adjacent spins in the 
bulk of the $XY$ model and its extended model, the extended Ising model \cite{Song,Song2}, and use it to 
detect the topological quantum phase transitions. In details, we find that the one-way deficit susceptibility 
of nearby two spins arrives its extremum 
value almost at the critical points of transverse field $XY$ model. Moreover, we investigated the one-way deficit 
in the extended Ising model and find that the one-way deficit is also able to characterize the topological 
phase transitions via its susceptibility. 
We also study the one-way deficit of thermal states of there models at nonzero temperatures.
Our results will not only enlighten extensive studies of the quantum information 
properties of ground states in different phases of critical systems, but also 
benefit a number of applications of these ground states, such as to detect the quantum phase 
transitions and to evaluate the capacity of quantum computations.

\section{One-way deficit in \emph{XY} model}\label{sec:FOC}
Like entanglement quantifications and other quantum correlations, the one way deficit considers a bipartite state $\rho^{ab}$
and is expressed as the difference of the von Neumann entropy before and after a on Neumann measurement
on one side. It is exactly given by \cite{streltsov}
\begin{eqnarray}
\Delta^{\rightarrow}(\rho^{ab})=\min\limits_{\{\Pi_{k}\}}S(\sum\limits_{k}\Pi_{k}\rho^{ab}\Pi_{k})-S(\rho^{ab}),\label{definition}
\end{eqnarray}
where $S(\cdot)$ denotes to the von Neumann entropy.
As a kind of quantum correlations besides entanglement and quantum discord, one-way deficit
links quantum thermodynamics with quantum correlations and deserve further investigations in
critical systems. 

We first consider the one-way deficit of ground states of the $XY$ model \cite{key-20} for the detection of the 
quantum phase transitions. The Hamiltonian of the $XY$ model is as follows \cite{key-21}:
\begin{equation}
H=-\sum_{j=1}^{L}\left(\frac{1+\gamma}{2}{\sigma}^{j}_{1}{\sigma}^{j+1}_{1}+\frac{1-\gamma}{2}{\sigma}^{j}_{2}{\sigma}^{j+1}_{2}+h{\sigma}^{j}_{3}\right)\label{ha}
\end{equation}
with $L$ the number of spins in the chain, ${\sigma}^{j}_{n}$ the $j$th spin Pauli operator in the 
direction $n=1,2,3$ and periodic boundary conditions assumed: $\sigma_n^{L+1}=\sigma_n^{1}$. 
The $XX$ model and transverse field Ising model thus correspond to the special cases for this general 
class of models: For the case that $\gamma\rightarrow0$, our model reduces to $XX$ model; and when 
$\gamma=1$, the model reduces to transverse field Ising model \cite{ising}. In fact, there exists
additional structure of interest in phase space beyond the breaking of phase flip symmetry at $h = 1$, 
which is the critical point between two quantum phases. It is worth noting that there exists a quarter 
of circle, $h^2 + \gamma^2 = 1$, on which the ground state is fully separable.

For the thermal ground states of $XY$ model (\ref{ha}), the Bloch representation of the reduced density 
matrix for two nearby spins at positions $i$ and $i+1$ has been obtained in \cite{Son} as
\begin{equation}\label{state11}
\rho^{ab}=\frac{1}{4}(I\otimes I+r{\sigma}_{3}\otimes I+sI\otimes {\sigma}_{3}+\sum_{n=1}^3c_n{\sigma}_n\otimes{\sigma}_n),
\end{equation}
where $I$ is the identity, $r=s=\langle{\sigma}^j_3\rangle$, $c_1=\langle{\sigma}^{j}_{1}{\sigma}^{j+1}_{1}\rangle$,
$c_2=\langle\sigma^{j}_{2}\sigma^{j+1}_{2}\rangle$, and $c_3=\langle{\sigma}^{j}_{3}{\sigma}^{j+1}_{3}\rangle.$
In the thermodynamic limit $L\gg1$ and at an approximately zero temperature $T\rightarrow0$, we have $\langle\sigma^{j}_{3}\rangle=G_0$, $\langle\sigma^{j}_{3}\sigma^{j+1}_{3}\rangle=\langle\sigma^{j}_{3}\rangle^{2}-G_{1}G_{-1}$, $\langle\sigma^{j}_{1}\sigma^{j+1}_{1}\rangle=G_{-1}$ and $\langle\sigma^{j}_{2}\sigma^{j+1}_{2}\rangle=G_{1}$, where 
\begin{equation}
G_{l}\equiv-\int_0^{\pi}\!\!\!\frac{d\phi}{\pi \omega_{\phi}}[\cos(l\phi)(h+\cos\phi)-\gamma\sin(l\phi)\sin\phi]
\end{equation}
and $\omega_{\phi}^2={(\gamma \sin\phi)^2+(h+\cos\phi)^2}$.

\begin{figure}[t]
\includegraphics[scale=0.55]{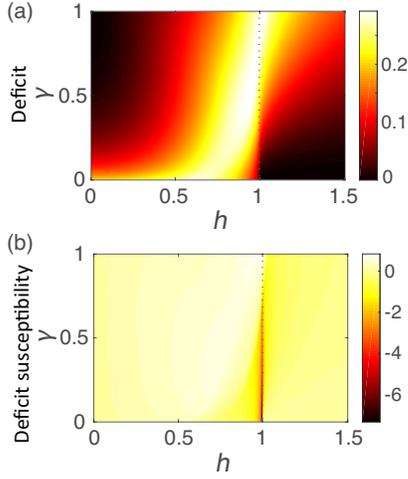}
\caption{(Color online) (a) One-way deficit and (b) deficit susceptibility of two adjacent spins in the bulk for the $XY$ model in the
thermodynamic limit as a function of the quantum parameter $h$ and $\gamma$. 
The dotted lines are for the critical points for the $XY$ models between different phases.}\label{fig:1}
\end{figure}

The eigenvalues of the $X$ states in Eq.~(\ref{state11}) are given by
\begin{eqnarray}
\lambda_{1,2}&=&(1-c_3\pm|c_{1}+c_{2}| )/4,\nonumber\\
\lambda_{3,4}&=&[1+c_3\pm\sqrt{(2r)^{2}+(c_{1}-c_{2})^{2}}]/4.
\end{eqnarray}
with which the von Neumann entropy is given by
\begin{eqnarray}\label{entropy3}
S(\rho^{ab})=-\sum\limits_{i=1}^{4}\lambda_{i}\log\lambda_{i}.
\end{eqnarray}
By Eqs.~(\ref{definition},\ref{entropy3}) and Eq.~(\ref{min3}) in the Appendix,
the one-way deficit of the $X$ states (\ref{state11}) is given by
\begin{equation}\label{deficit}
\Delta^{\rightarrow}(\rho^{ab})
=-\!\!\!\!\!\!\min\limits_{\{z_{1}^{2}+z_{2}^{2}+z_{3}^{2}=1\}}\sum\limits_{i=1}^{4}w_{i}\log w_{i}+\sum\limits_{i=1}^{4}\lambda_{i}\log\lambda_{i},
\end{equation}
where we have
\begin{align}
&w_{1,2}
=\frac{1}{4}\left[1-sz_{3}\pm\sqrt{(r-c_{3}z_{3})^2+c_{1}^{2}z_{1}^{2}+c_{2}^{2}z_{2}^{2}}\right],\label{e1}\\
&w_{3,4}
=\frac{1}{4}\left[1+sz_{3}\pm\sqrt{(r+c_{3}z_{3})^2+c_{1}^{2}z_{1}^{2}+c_{2}^{2}z_{2}^{2}}\right].\label{e2}
\end{align}
with a constraint condition as $z_{1}^{2}+z_{2}^{2}+z_{3}^{2}=1$.

\begin{figure}[t]
\includegraphics[scale=0.52]{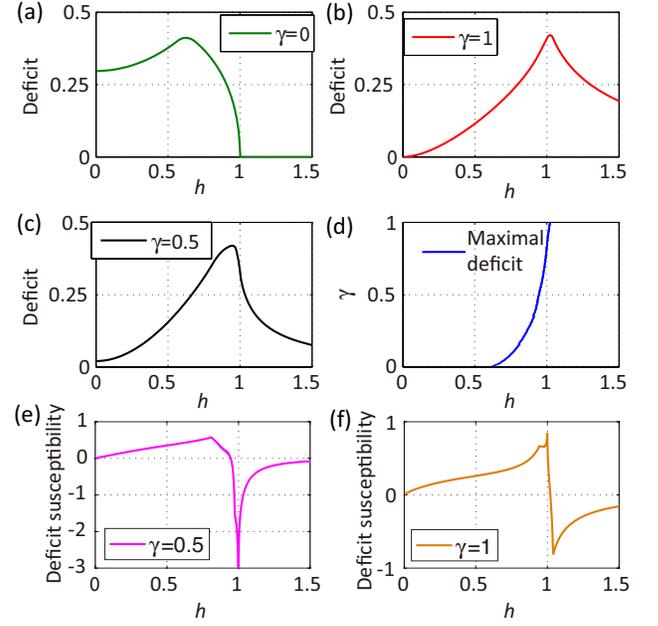}
\caption{(Color online) One-way deficit of two adjacent spins in the bulk for the $XY$ model:
(a-c) One-way deficit of reduced nearby two-qubit ground states of $XY$ models 
for $\gamma\rightarrow0$ ($XX$ model), $\gamma=1$ (transverse field Ising model) 
and $\gamma=0.5$ respectively. (d) Maximal one-way deficit of the $XY$ model.
(e,f) Deficit susceptibility of reduced nearby two-qubit ground states of $XY$ models for 
$\gamma=0.5$ and $\gamma=1$, respectively.}\label{fig:2}
\end{figure}

Then, we use the equations above to calculate the one-way deficit of two adjacent spins in 
the bulk for the $XY$ model and analyze the one-way deficit and one-way deficit susceptibility
that is defined as $\chi=\partial \Delta^{\rightarrow}(\rho^{ab})/\partial h$. The main results 
are shown in Figs.~\ref{fig:1} and \ref{fig:2}, in which we plot the one-way deficit and the one-way deficit 
susceptibility of two adjacent spins in the bulk of $XY$ model as a function of $h$ and $\gamma$. 
Given $\gamma$ a fixed value, we observe that as the transverse field strength $h$ increases,
the one-way deficit increases for small $h$ and decreases for large $h$,  see Fig.~\ref{fig:2}(c) for 
$\gamma=0.5$. When $\gamma\rightarrow1$, the model reduces to the Ising model and the 
maximum of the one-way deficit is attained near $h=1$, see Figs.~\ref{fig:2}(b,d,f). From Figs.~\ref{fig:1}(b) 
and \ref{fig:2}(e,f), we clearly show that the one-way deficit susceptibility reach its extremum when 
the quantum phase transitions occur.

In details, for the case that $\gamma\rightarrow0$ as shown in Fig.~\ref{fig:2}(a), the $XY$ model reduces
into the $XX$ model, where we find that the one-way deficit is nonzero in the domain $h\in[0,1)$ and 
then suddenly becomes zero as $h\geq1$. As the $XX$ model undergoes a first order transition at the 
critical point $h=1$ from fully polarized to a critical phase with quasi-long-range order, we conclude 
that one-way deficit can effectively detect quantum phase of the $XX$ model. The conclusion is in 
consistent with the result obtained in \cite{wang}. In Fig.~\ref{fig:2}(b) for the case that $\gamma=1$, the model 
reduces to transverse field Ising model and we find that one-way deficit of the Ising model increases for small 
$h$ and decreases for large $h$. When one-way deficit susceptibility reach its extremum nearly at 
$h=1$, transverse field Ising model undergoes a first order transition. Generally, as shown in Fig.~\ref{fig:1}(b),
we infer that one-way deficit  can be used to detect quantum phase of the $XY$ model given different values 
of $\gamma$. 

\begin{figure}[t]
\includegraphics[scale=0.45]{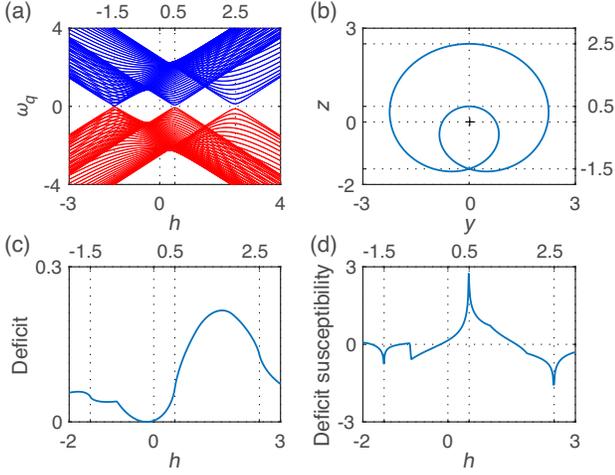}
\caption{(Color online) (a) Energy spectra given $L=200$ sites, (b) trajectory of winding 
vector, (c) one-way deficit and (d) its susceptibility of thermal ground states as a function of $h$
for the extended Ising model with parameters $\gamma=1$, $\delta=1$ and $\lambda=3/2$.}\label{fig:3}
\end{figure}

\section{One-way deficit in extended Ising model}\label{sec:FOC2}
In recent years, the quantum topological order \cite{Wen1} in extended critical systems \cite{Song,Song2,Niu,LRKitaev}
has become more and more important in both topological quantum computation and 
condensed matter physics \cite{Wen2,te1,te2,es,Kitaev,Xiang, Meng}.
Here we consider the extended Ising model that contains several kinds of topological phases and 
is written as \cite{Song,Song2}
\begin{align}
&H=\sum_{j=1}^{L}\lambda\sigma^j_3\left(\frac{1+\delta}{2}{\sigma}^{j-1}_{1}{\sigma}^{j+1}_{1}+\frac{1-\delta}{2}{\sigma}^{j-1}_{2}{\sigma}^{j+1}_{2}\right)\nonumber\\
&~~~~+\sum_{j=1}^L\left(\frac{1+\gamma}{2}{\sigma}_{1}^{j}{\sigma}^{j+1}_{1}+\frac{1-\gamma}{2}{\sigma}^{j}_{2}{\sigma}^{j+1}_{2}+h{\sigma}^{j}_{3}\right)\label{ha1}
\end{align}
with the periodic boundary conditions assumed $\sigma_n^0=\sigma_n^L$ and 
$\sigma_n^{L+1}=\sigma_n^1$. The Bloch representation of the reduced density matrix
for two nearby spins at positions $j$ and $j+1$ is shown in Eq.~(\ref{state11})
with parameters $\tilde{r}=\tilde{s}=\tilde{G}_0$, $\tilde{c}_1=\tilde{G}_{-1}$,
$\tilde{c}_2=\tilde{G}_1$, and $\tilde{c}_3=\tilde{G}_0^2-\tilde{G}_1\tilde{G}_{-1}$,
where the spin-spin correlation functions with $\beta=1/T$ the inverse of temperature can be written as 
\begin{eqnarray}
\tilde{G}_{l}\equiv-\int_{0}^{\pi}\!\!\frac{d\phi\tanh(\beta\tilde{\omega}_\phi)}{\pi }\cos(l\phi-\tilde{\Theta}_{\phi})%\\
%\times[\cos(r\phi)(h+\cos\phi+\lambda\cos2\phi)-\sin(r\phi)(\gamma\sin\phi+\lambda\delta\sin2\phi)].
\end{eqnarray}
with
\begin{align}
&\tilde{\Theta}_\phi=\arctan\frac{\gamma \sin\phi+\lambda\delta \sin2\phi}{h+\cos\phi+\lambda\cos2\phi},\\
&\tilde{\omega}_{\phi}^2={(\gamma \sin\phi+\lambda\delta \sin2\phi)^2+(h+\cos\phi+\lambda\cos2\phi)^2}.%{h^2}
\end{align}
Similarly, we can calculate the one-way deficit of the states (\ref{state11}) of two nearby spins in the 
extended Ising model using Eq.~(\ref{deficit}). Furthermore, compared with the results of quantum discord 
in Ref.~\cite{xzzhang}, we should emphasize that the one-way deficit is proved to be larger than the quantum 
discord \cite{blye} for the $X$ states shown in Eq.~(\ref{state11})  and shows more properties and structures 
of the ground states of the extended Ising models.

\begin{figure}[t]
\includegraphics[scale=0.45]{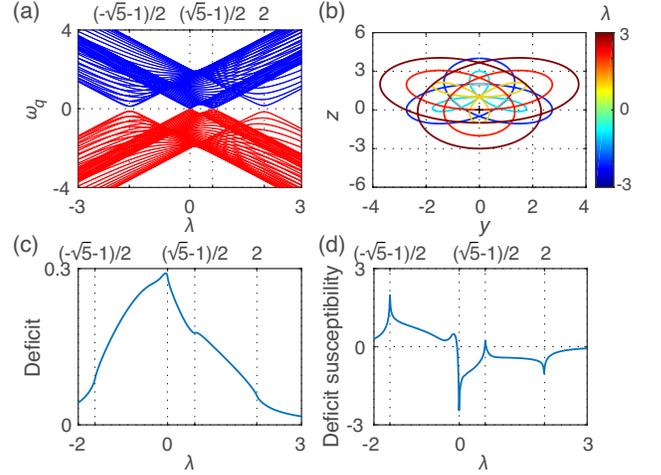}
\caption{(Color online) (a) Energy spectra given $L=200$ sites, (b) trajectories of winding 
vectors for different value of $\lambda$, (c) one-way deficit and (d) its susceptibility of 
thermal ground states as a function of $\lambda$
for the extended Ising model with parameters $\gamma=1$, $\delta=-1$ and $h=1$.}\label{fig:4}
\end{figure}

For the critical point between two topological phases with different winding numbers at zero temperature,
one needs to solve the characteristic equation $g(\zeta)=0$ with zeros on the contour $|\zeta|=1$ in the 
complex plane, where the complex characteristic function $g(\zeta)$ is defined in the Appendix B and carefully 
introduced and discussed in Ref.~\cite{yrzhang}. For instance, we set the parameters of extended Ising 
model as $\gamma=1$, $\delta=1$, $\lambda=3/2$ and change the value of $h$, and the characteristic 
equation is written as
\cite{yrzhang}
\begin{eqnarray}
g(\zeta)=3\zeta^2/2+\zeta-h=0,
\end{eqnarray}
with which the critical points for the emergence of topological phase transitions at 
$h=2.5$ for $\zeta=1$, $h=0.5$ for $\zeta=-1$ and $h=-1.5$ for $\zeta=\exp[\pm i\arccos(-1/3)]$
can be calculated \cite{yrzhang}. For this example, energy spectra for $L = 200$ sites and the 
trajectory of winding vector are shown in Fig.~\ref{fig:3}(a,b), respectively. 
In Fig.~\ref{fig:3}(c,d), one-way deficit and its susceptibility as functions of $h$ are plotted. 
It is shown that one-way deficit susceptibility reaches its extremum nearly at the critical points of 
topological phase transitions.

Moreover, we consider the parameters of extended Ising model as $\gamma=1$, $\delta=-1$, $h=1$
and change the value of $\lambda$ at zero temperature. We can obtain the critical points of topological 
phase transitions by solving the characteristic equation: \cite{yrzhang}
\begin{eqnarray}
g(\zeta)=\lambda\zeta^2+\zeta^{-1}-1=0
\end{eqnarray}
where we can obtain the critical points at $\lambda=0$ for $\zeta=1$,
$\lambda=2$ for $\zeta=-1$, $\lambda=(-\sqrt{5}-1)/2$ for $\zeta=\exp\{\pm i\arccos[(1-\sqrt{5})/4]\}$, and
$\lambda=(\sqrt{5}-1)/2$ for $\zeta=\exp\{\pm i\arccos[(1+\sqrt{5})/4]\}$.
Similar results of one-way deficit are shown in Fig.~\ref{fig:4}. 

\begin{figure}[t]
\includegraphics[scale=0.43]{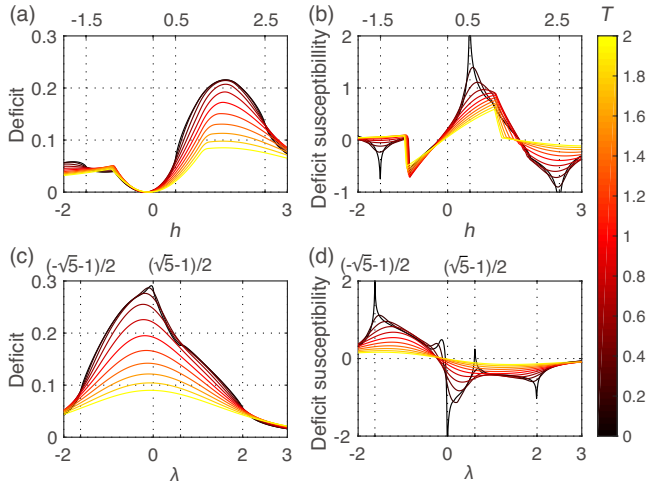}
\caption{(Color online) One-way deficit and its susceptibility of thermal states of the extended Ising models with 
parameters (a,b) $\gamma=1$, $\delta=1$, $\lambda=3/2$ and (c,d) $\gamma=1$, $\delta=-1$ and $h=1$ at 
different temperatures $T$.}\label{fig:5}
\end{figure}

Therefore, we can conclude 
that the distinct critical behaviors of one-way deficit, presented by the one-way deficit susceptibility, 
effectively characterize the topological quantum phase transitions of the extended Ising model.
For completeness, we consider impact of the noise at nonzero temperature, and show one-way deficit and its 
susceptibility of thermal states at different temperatures in Fig.~\ref{fig:5}. It is shown that the extremum points of 
the one-way deficit susceptibility shift as the temperature increases, and the detection of topological 
phase transition points could be accurate at a low temperature $T\ll1$.

\section{conclusion}\label{sec:conclusion}
In this paper, we present a method to evaluate the one-way deficit of the thermal states of
two adjacent spins in the bulk for the $XY$ model and the extended Ising model in the 
thermodynamic limit. The diagram of the one-way deficit and deficit susceptibility of the 
ground states of $XY$ model are plotted. We find it effective to use one-way deficit to detect 
 quantum phase transitions of the $XY$ model given different values of the parameter $\gamma$. 
 Moreover, we show that the distinct critical behaviors of one-way deficit, presented by the one-way 
 deficit susceptibility, effectively characterize the topological quantum phase transitions of the 
 extended Ising model. On one hand, our results may shed lights on the study of properties of 
 quantum correlations in different quantum phases of many body systems. On the other hand, 
 our investigations will also benefit a number of applications in quantum physics including the 
 detection of topological orders and the evaluation of the capacity of quantum computation in 
 critical systems. Also numerical techniques such as DMRG, MPS and exact diagonalization 
 methods deserve to be used to investigate   extensive problems of quantum topological phase transitions 
 from the perspective of quantum information and quantum correlations, including finite-temperature 
 phase transitions, dynamical phase transitions of more quantum many-body models.

\begin{acknowledgments}
We would like to thank Jin-Jun Chen for useful discussions.
 This work was supported by the Science and Technology Research Plan Project of the Department of Education of Jilin Province in the Twelfth Five-Year Plan.
 Ministry of Science and
Technology of China (Grants No. 2016YFA0302104 and
2016YFA0300600), National Natural Science Foundation
of China (Grant No. 91536108) and Chinese Academy of
Sciences (Grants No. XDB01010000 and XDB21030300).
 %, the National Natural Science Foundation of China under grant Nos. 11175248, 11275131, 11305105.
\end{acknowledgments}

\appendix
\section{One-way deficit of $X$ states}
In this section, we evaluate the one-way deficit of $X$ states in Eq.~(\ref{state11}). 
Let $\{\Pi_{k}=|k\rangle\langle k|, k=0, 1\}$
be the local measurement on the party $b$ along the computational base ${|k\rangle}$;  then any von Neumann measurement for the party $b$ can be written as
\begin{eqnarray}\label{bk}
\{B_{k}=V\Pi_{k}V^{\dag}: k=0, 1\}
\end{eqnarray}
given some unitary operator $V\in U(2)$. For any  $V$,
\begin{eqnarray}\label{v}
V=tI+i\vec{y}\cdot\vec{\sigma}=\left(
                                 \begin{array}{cc}
                                   t+y_{3}i & y_{2}+y_{1}i \\
                                   -y_{2}+y_{1}i & t-y_{3}i \\
                                 \end{array}
                               \right).
\end{eqnarray}
with $t\in \mathbb{R}$, $\vec{y}=(y_{1}, y_{2}, y_{3})\in \mathbb{R}^{3}$, and
\begin{eqnarray}\label{ty}
t^{2}+y_{1}^{2}+y_{2}^{2}+y_{3}^{2}=1,
\end{eqnarray}
after the measurement ${B_{k}}$, the state $\rho^{ab}$ will be changed into the ensemble $\{{\rho_{k}, p_{k}}\}$ with
\begin{eqnarray}\label{pp}
\rho_{k}&=&\frac{1}{p_{k}}(I\otimes B_{k})\rho(I\otimes B_{k}),\\
p_{k}&=&\textrm{Tr}(I\otimes B_{k})\rho(I\otimes B_{k}).
\end{eqnarray}
To evaluate $\rho_{k}$ and $p_{k}$, we write
\begin{eqnarray}
p_{k}\rho_{k}&=&(I\otimes B_{k})\rho(I\otimes B_{k})\nonumber\\
&=&\frac{1}{4}(I\otimes V)(I\otimes \Pi_{k})[I+r\sigma_{3}\otimes I+sI\otimes V^{\dag}\sigma_{3}V^{\dag}\nonumber\\
& &+\sum_{j=1}^3 c_{j}\sigma_{j}\otimes (V^{\dag} \sigma_{j} V)](I\otimes \Pi_{k})(I\otimes V^{\dag}).
\end{eqnarray}
Using the relations \cite{luo}
\begin{eqnarray}
V^{\dag}\sigma_{1}V&=&(t^{2}+y_{1}^{2}-y_{2}^{2}-y_{3}^{2})\sigma_{1}+2(ty_{3}+y_{1}y_{2})\sigma_{2}\nonumber\\
& &+2(-ty_{2}+y_{1}y_{3})\sigma_{3},\label{condition2}  \\
V^{\dag}\sigma_{2}V&=&2(-ty_{3}+y_{1}y_{2})\sigma_{1}+(t^{2}+y_{2}^{2}-y_{1}^{2}-y_{3}^{2})\sigma_{2}\nonumber\\
& &+2(ty_{1}+y_{2}y_{3})\sigma_{3},\label{condition3} \\
V^{\dag}\sigma_{3}V&=&2(ty_{2}+y_{1}y_{3})\sigma_{1}+2(-ty_{1}+y_{2}y_{3})\sigma_{2}\nonumber\\
& &+(t^{2}+y_{3}^{2}-y_{1}^{2}-y_{2}^{2})\sigma_{3},\label{condition4}
\end{eqnarray}
and
\begin{eqnarray}\label{condition5}
\Pi_{0}\sigma_{3}\Pi_{0}=\Pi_{0}, \Pi_{1}\sigma_{3}\Pi_{1}=-\Pi_{1},\Pi_{j}\sigma_{k}\Pi_{j}=0,
\end{eqnarray}
for $j=0, 1, k=1, 2$,
we obtain
\begin{eqnarray}
p_{0}\rho_{0}&=&\frac{1}{4}[I+sz_{3}I+c_{1}z_{1}\sigma_{1}+c_{2}z_{2}\sigma_{2}+(r+c_{3}z_{3})\sigma_{3}]\nonumber\\
& &\otimes(V\Pi_{0}V^{\dag}),\\
p_{1}\rho_{1}&=&\frac{1}{4}[I-sz_{3}I-c_{1}z_{1}\sigma_{1}-c_{2}z_{2}\sigma_{2}+(r-c_{3}z_{3})\sigma_{3}]\nonumber\\
& &\otimes(V\Pi_{1}V^{\dag}),
\end{eqnarray}
where
\begin{eqnarray}
z_{1}=2(-ty_{2}+y_{1}y_{3}),\
z_{2}=2(ty_{1}+y_{2}y_{3}), \\
z_{3}=t^{2}+y_{3}^{2}-y_{1}^{2}-y_{2}^{2}.\ \ \ \ \ \ \ \ \ \ \ \ \ \label{condition6}
\end{eqnarray}
Then, we will evaluate the eigenvalues of $\sum_{k}\Pi_{k}\rho^{ab}\Pi_{k}$ by
$\sum_{k}\Pi_{k}\rho^{ab}\Pi_{k}=p_{0}\rho_{0}+p_{1}\rho_{1}$, and
\begin{align}
p_{0}\rho_{0}+p_{1}\rho_{1}
%&=&\frac{1}{4}[(I+r\sigma_{3})+(rz_{3}I+cz_{1}\sigma_{1}+cz_{2}\sigma_{2}+c_{3}z_{3}\sigma_{3})]\otimes(V\Pi_{0}V^{\dag})\nonumber\\
%& &+\frac{1}{4}[(I+r\sigma_{3})-(rz_{3}I+cz_{1}\sigma_{1}+cz_{2}\sigma_{2}+c_{3}z_{3}\sigma_{3})]\otimes(V\Pi_{1}V^{\dag})\nonumber\\
%&=&\frac{1}{4}(I+r\sigma_{3})\otimes(V\Pi_{0}V^{\dag}+V\Pi_{1}V^{\dag})\nonumber\\
%& &+\frac{1}{4}(rz_{3}I+cz_{1}\sigma_{1}+cz_{2}\sigma_{2}+c_{3}z_{3}\sigma_{3})
%\otimes(V\Pi_{0}V^{\dag}-V\Pi_{1}V^{\dag})\nonumber\\
=\frac{1}{4}(I+r\sigma_{3})\otimes I\ \ \ \ \ \ \ \ \ \ \ \ \ \ \ \ \ \ \ \ \ \ \ \ \ \ \ \ \ \ \ \nonumber\\
+\frac{1}{4}(sz_{3}I+c_{1}z_{1}\sigma_{1}+c_{2}z_{2}\sigma_{2}+c_{3}z_{3}\sigma_{3})
\otimes V\sigma_{3}V^{\dag}.
\end{align}
The eigenvalues of $p_{0}\rho_{0}+p_{1}\rho_{1}$  are the same with the eigenvalues of the states $(I\otimes V^{\dag})(p_{0}\rho_{0}+p_{1}\rho_{1})(I\otimes V)$, and
\begin{align}\label{value2}
& (I\otimes V^{\dag})(p_{0}\rho_{0}+p_{1}\rho_{1})(I\otimes V)\nonumber\\
&=\frac{1}{4}(sz_{3}I+c_{1}z_{1}\sigma_{1}+c_{2}z_{2}\sigma_{2}+c_{3}z_{3}\sigma_{3})\otimes\sigma_{3}\nonumber\\
& +\frac{1}{4}(I+r\sigma_{3})\otimes I.
\end{align}
 The eigenvalues of the states in Eq.~(\ref{value2}) are given in Eqs.~(\ref{e1},\ref{e2}).
Thus, the entropy of $\sum_{k}\Pi_{k}\rho^{ab}\Pi_{k}$ is 
\begin{equation}
S(\sum_{k}\Pi_{k}\rho^{ab}\Pi_{k})=-\sum_{i=1}^{4}w_{i}\log w_{i}. 
\end{equation}
When $\gamma, h$ are fixed,
$r, s, c_{1}, c_{2},c_{3}$ are constant. By using  $z_{1}^{2}+z_{2}^{2}+z_{3}^{2}=1$,
it converts the problem about $\min_{\{\Pi_{k}\}} S(\sum_{k}\Pi_{k}\rho^{ab}\Pi_{k})$ to the problem about the function of three variables $z_{1},z_{2},z_{3}$ for minimum, that is
\begin{equation}\label{min3}
\min_{\{\Pi_{k}\}} S(\sum\limits_{k}\Pi_{k}\rho^{ab}\Pi_{k})=\!\!\!\!\!\!\!\min \limits_{\{z_{1}^{2}+z_{2}^{2}+z_{3}^{2}=1\}}\!\!\!\!\!\!\!S(\sum\limits_{k}\Pi_{k}\rho^{ab}\Pi_{k}).
\end{equation}
Therefore, by Eqs.~(\ref{definition}), (\ref{entropy3}), (\ref{min3}), the one-way deficit of $X$ states in Eq. (\ref{state11}) is obtained as
shown in Eq.~(\ref{deficit}).

\section{Diagonalization, winding numbers and characteristic complex functions of the extended Ising model,} 
The Hamiltonian of extended Ising model (\ref{ha1}) can be mapped to a spinless fermion Hamiltonian 
by the Jordan-Wigner transformation $-c_1=\sigma_1^+=(\sigma^x_1+i\sigma^y_1)/2$, 
$-c_j=\sigma_j^+\prod_{i=1}^{j-1}\sigma^z_i$ \cite{Song}.
In the thermodynamic limit $L\gg 1$, we can use the Bogoliubov-Fourier transformation 
to rewrite a Bogoliubov-de Gennes (BdG) Hamiltonian as \cite{Song}
\begin{equation}
H=\sum_{\phi}(c_\phi^\dag\ c_{-\phi})\mathcal{H}_\phi\left(\begin{array}{c}{c_\phi}\\{c_{-\phi}^\dag}\end{array}\right),
\end{equation} where the complete set of 
wavevectors is $\phi=2\pi m/L$ with $m=-\frac{L-1}{2},-\frac{L-3}{2},\cdots,\frac{L-3}{2},\frac{L-1}{2}$. 
Here, we can write \cite{Song}
\begin{equation}
\mathcal{H}_\phi=\bm{r}(\phi)\cdot \bm{\sigma}
\end{equation} 
with the vector $\bm{r}(\phi)=(0~Y(\phi)~Z(\phi))$ in the auxiliary two-dimensional $y-z$ space, 
\begin{align}
&Y(\phi)=\lambda\delta\sin(2\phi)+\gamma\sin\phi,\\ 
&Z(\phi)=\lambda\cos(2\phi)+\cos\phi-h
\end{align}
and $\bm{\sigma}=(\sigma_1~\sigma_2~\sigma_3)$.
The winding number of the closed loop in auxiliary $y-z$ 
plane around the origin point can be written as 
\begin{equation}
\nu=\frac{1}{2\pi}\oint(YdZ-ZdY)/|\bm{r}|^2,
\end{equation}
which is used to identify different topological orders in 
the BDI class one-dimensional fermion systems \cite{revi}.

Using the Bogoliubov transformation $c_\phi=\cos\frac{\Theta}{2}\eta_\phi+i\sin\frac{\Theta}{2}\eta_{-\phi}^\dag$ 
with $\tan\Theta\equiv Y(\phi)/Z(\phi)$, we can diagonalize the Hamiltonian as 
\begin{equation}
{H}=\sum_\phi\omega_\phi(\eta_\phi^\dag\eta_\phi-{1}/{2})
\end{equation} 
and obtain the ground state as
\begin{eqnarray}
|\mathcal{G}\rangle=\prod_{\phi}[\cos\frac{\Theta}{2}+i\sin\frac{\Theta}{2}\eta_\phi^{\dag}\eta_{-\phi}^\dag]|0\rangle,
\end{eqnarray}
where $|0\rangle$ is the vacuum state and the energy spectra are 
\begin{eqnarray}
\omega_\phi=\sqrt{Y(\phi)^2+Z(\phi)^2}.
\end{eqnarray}
Via a substitute $\zeta(\phi)\equiv\exp(i\phi)$, the characteristic function is defined as \cite{yrzhang}
\begin{align}
&g(\zeta)\equiv Z(\phi)+iY(\phi)\\
&~~=\lambda[\zeta^2+(1-\delta)\zeta^{-2}/2]+\zeta+(1-\gamma)\zeta^{-1}/2-h
\end{align}
with which we can calculate the critical points for the quantum topological phase transitions by the
characteristic equation $g(\zeta)=0$ with $|\zeta|=1$ required.

\end{document}